\documentclass[useAMS,usenatbib]{mn2e}

\usepackage{latexsym,graphicx,natbib}

\usepackage{color}

\newcommand\kms{{\rm\,km\,s^{-1}}}

\newcommand\lsun{\rm\,L_\odot}
\newcommand\hii{H\,{\sc ii} \,}

\newcommand{\MC}{\multicolumn}

\def\apgt{\ {\raise-.5ex\hbox{$\buildrel>\over\sim$}}\ }
\def\aplt{\ {\raise-.5ex\hbox{$\buildrel<\over\sim$}}\ }

\title[MN48: a new Galactic bona fide LBV]{MN48: a new Galactic bona fide luminous blue variable revealed by {\it Spitzer} and
SALT\footnotemark[0]\thanks{Based on observations obtained with
the Southern African Large Large Telescope (SALT), programmes
\mbox{2010-1-RSA\_OTH-001}, \mbox{2013-1-RSA\_OTH-014} and
\mbox{2013-2-RSA\_OTH-003} (PI: A.Y.\,Kniazev).}}
\author[A.Y Kniazev et al.]
       {A. Y.~Kniazev,$^{1,2,3}$\thanks{E-mail: akniazev@saao.ac.za}
    V. V.~Gvaramadze,$^{3,4,5}$
    L. N.~Berdnikov$^{3,5,6}$ \\
    $^{1}$South African Astronomical Observatory, PO Box 9, 7935 Observatory, Cape Town, South Africa \\
    $^{2}$Southern African Large Telescope Foundation, PO Box 9, 7935 Observatory, Cape Town, South Africa \\
    $^{3}$Sternberg Astronomical Institute, Lomonosov Moscow State University, Universitetskij Pr. 13, Moscow 119992, Russia\\
    $^{4}$Space Research Institute, Russian Academy of Sciences, Profsoyuznaya 84/32, Moscow 117997, Russia \\
    $^{5}$Isaac Newton Institute of Chile, Moscow Branch, Universitetskij Pr. 13, Moscow 119992, Russia \\
    $^{6}$Astronomy and Astrophysics Research division,  Entoto Observatory and Research Center, PO Box 8412 Addis Ababa, Ethiopia \\
    }
\begin{document}

\date{Accepted 2016 April 13. Received 2016 April 13; in original form 2016 February 7}


\maketitle

\label{firstpage}

\begin{abstract}
In this paper, we report the results of spectroscopic and
photometric observations of the candidate evolved massive star
MN48 disclosed via detection of a mid-infrared circular shell
around it with the {\it Spitzer Space Telescope}. Follow-up
optical spectroscopy of MN48 with the Southern African Large
Telescope (SALT) carried out in 2011--2015 revealed significant
changes in the spectrum of this star, which are typical of
luminous blue variables (LBVs). The LBV status of MN48 was further
supported by photometric monitoring which shows that in 2009--2011
this star has brightened by $\approx$0.9 and 1 mag in the $V$ and
$I_{\rm c}$ bands, respectively, then faded by $\approx$1.1 and
1.6 mag during the next four years, and apparently started to
brighten again recently. The detected changes in the spectrum and
brightness of MN48 make this star the 18th known Galactic bona
fide LBV and increase the percentage of LBVs associated with
circumstellar nebulae to more than 70 per cent. We discuss the
possible birth place of MN48 and suggest that this star might have
been ejected either from a putative star cluster embedded in the
\hii region IRAS\,16455$-$4531 or the young massive star cluster
Westerlund\,1.
\end{abstract}


\section{Introduction}
\label{sec:intro}

\begin{figure*}
\begin{center}
\includegraphics[width=17cm,angle=0]{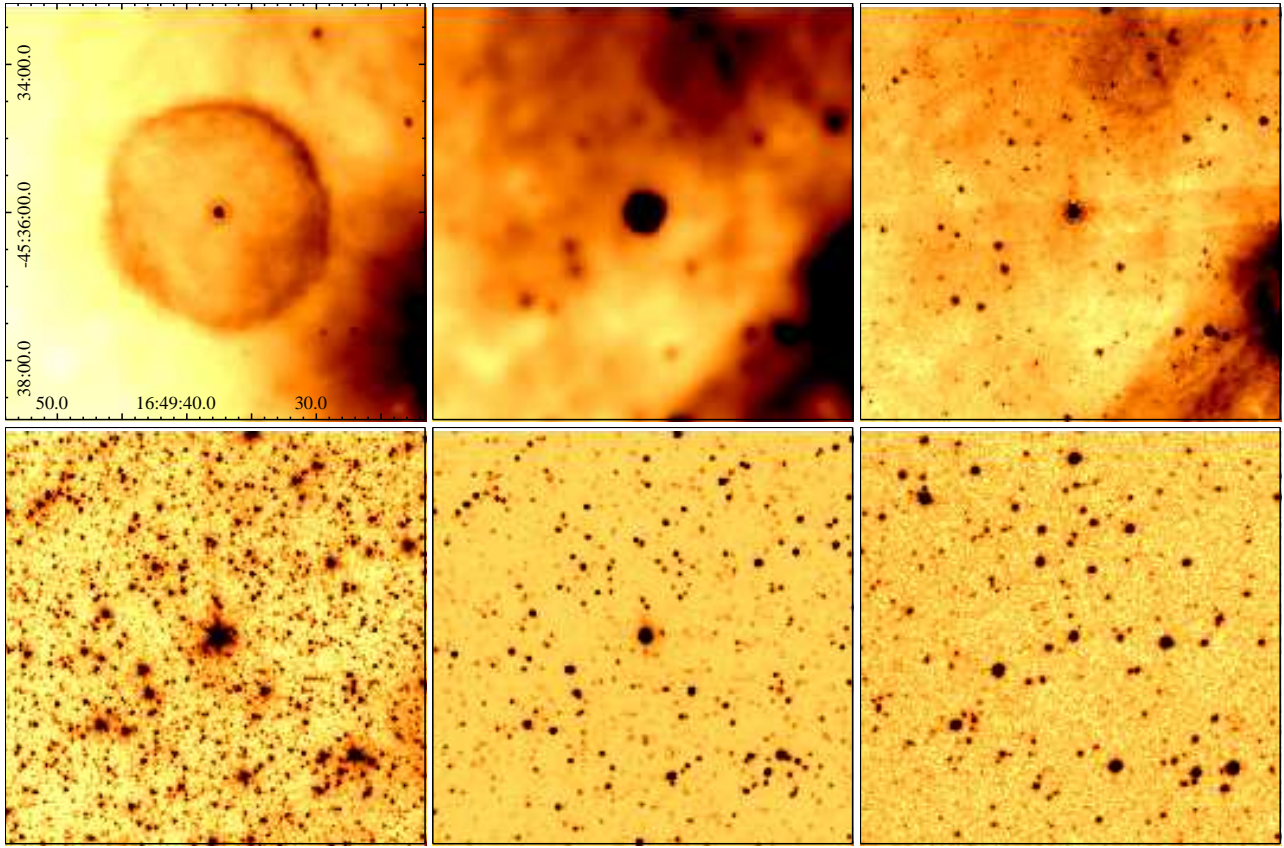}
\end{center}
\caption{From left to right, and from top to bottom: {\it Spitzer}
MIPS 24\,$\mu$m, {\it WISE} 12\,$\mu$m,  IRAC 8 and 3.6\,$\mu$m,
2MASS $K_{\rm s}$ band and DSS II red band images of the region
containing the nebula MN48 and its central star (the scale of the
images is the same). A bright emission to the south-west of MN48
is the \hii region IRAS\,16455$-$4531. The coordinates are in
units of RA (J2000) and Dec. (J2000) on the horizontal and
vertical scales, respectively.} \label{fig:neb}
\end{figure*}

Luminous blue variables (LBVs) belong to a very rare class of
massive stars at advanced evolutionary stages (Conti 1984),
characterized by spectacular photometric and spectral variability
(Humphreys \& Davidson 1994; van Genderen 2001). During major
eruptions, LBVs can be confused with supernovae (e.g. Goodrich et
al. 1989; Filippenko et al. 1995). The nature of the LBV
phenomenon as well as the evolutionary status of LBVs remain
unclear. Although it is believed that the LBV phase is
intermediate between the main sequence and Wolf-Rayet phases (e.g.
Langer et al. 1994; Stothers \& Chin 1996), there are strong
indications that some LBVs are immediate precursors of supernovae
(e.g. Kotak \& Vink 2006; Gal-Yam \& Leonard 2009; Groh, Meynet \&
Ekstr\"{o}m 2013; Groh et al. 2013). Revealing of new examples of
LBVs might be of crucial importance for understanding these
enigmatic objects.

The majority ($\approx$70 per cent; Kniazev, Gvaramadze \&
Berdnikov 2015) of the LBVs are surrounded by compact (circular or
bipolar) circumstellar nebulae, composed of stellar material
ejected during the LBV and/or preceding evolutionary stages (Nota
et al. 1995). Detection of such nebulae serves as an indication
that their underlying stars are LBVs or related evolved massive
stars. Subsequent searches for significant changes in the
brightness and spectra of these stars are necessary to determine
which of them are bona fide LBVs.

Until recently, only three LBV candidates were revealed through
the detection of (mid-infrared) circumstellar shells and follow-up
spectroscopy of their central stars (Waters et al. 1996; Egan et
al. 2002; Clark et al. 2003). This is mostly because of low
angular resolution and sensitivity of the data (obtained with the
{\it InfraRed Astronomical Satellite} and the {\it Midcourse Space
Experiment} satellite) being used. The situation has been improved
drastically with the advent of the {\it Spitzer Space Telescope}.
Searches for compact nebulae using the {\it Spitzer} data resulted
in discovery of many dozens of compact nebulae, whose geometry is
reminiscent of that of the nebulae associated with the already
known LBVs (Gvaramadze, Kniazev \& Fabrika 2010b; Wachter et al.
2010; Mizuno et al. 2010). Spectroscopy of central stars of these
nebulae showed that many of them have spectra similar to those of
the LBVs, which further supports the LBV nature of these stars
(Gvaramadze et al. 2010a,b, 2015a; Wachter et al. 2010, 2011;
Stringfellow et al. 2012a,b; Flagey et al. 2014; Kniazev \&
Gvaramadze 2016).

The majority of the compact nebulae discovered with {\it Spitzer}
were detected in $24 \, \mu$m images obtained with the Multiband
Imaging Photometer for {\it Spitzer} (MIPS; Rieke et al. 2004)
within the framework of the 24 and 70 Micron Survey of the Inner
Galactic Disk with MIPS (MIPSGAL; Carey et al. 2009) and other
{\it Spitzer}
programmes\footnote{http://irsa.ipac.caltech.edu/Missions/spitzer.html}.
Some of them have obvious counterparts at shorter wavelengths and
can be detected at 8, 5.8, 4.5 and $3.6\,\mu$m images obtained
with the {\it Spitzer} Infrared Array Camera (IRAC; Fazio et al.
2004). In some cases, the nebulae were discovered using the IRAC
data only because the corresponding MIPS images are oversaturated
(Gvaramadze et al. 2010b). The number of new detections of compact
nebulae continues to increase (Gvaramadze et al. 2011, 2012a;
Kniazev \& Gvaramadze 2016) with the release of an all-sky survey
in 3.4, 4.6, 12 and 22\,$\mu$m bands performed with the {\it
Wide-field Infrared Survey Explorer} ({\it WISE}; Wright et al.
2010). This, in turn, leads to further increase of the number of
newly identified LBV candidates (Gvaramadze et al. 2012a; Kniazev
\& Gvaramadze 2016).

In 2009 we started a monitoring campaign of LBV candidates
revealed with {\it Spitzer} (and later on with {\it WISE}) in
order to search for their possible spectral and photometric
variability, and potentially to prove that at least some of them
are bona fide LBVs. Since then, we have confirmed the LBV status
of three of these stars: Wray\,16-137 (Gvaramadze et al. 2014),
WS1 (Kniazev, Gvaramadze \& Berdnikov 2015) and MN44 (Gvaramadze,
Kniazev \& Berdnikov 2015b). In this paper, we present
observations indicating that a fourth star in our sample, MN48, is
a bona fide LBV as well. In Section\,\ref{sec:neb}, we present the
star and its nebula. In Section\,\ref{sec:obs}, we describe our
spectroscopic and photometric observations. The results are
discussed in Section\,\ref{sec:dis}.

\section{MN48 and its mid-infrared nebula}
\label{sec:neb}

\begin{figure*}
\begin{center}
\includegraphics[angle=270,width=1.0\textwidth,clip=]{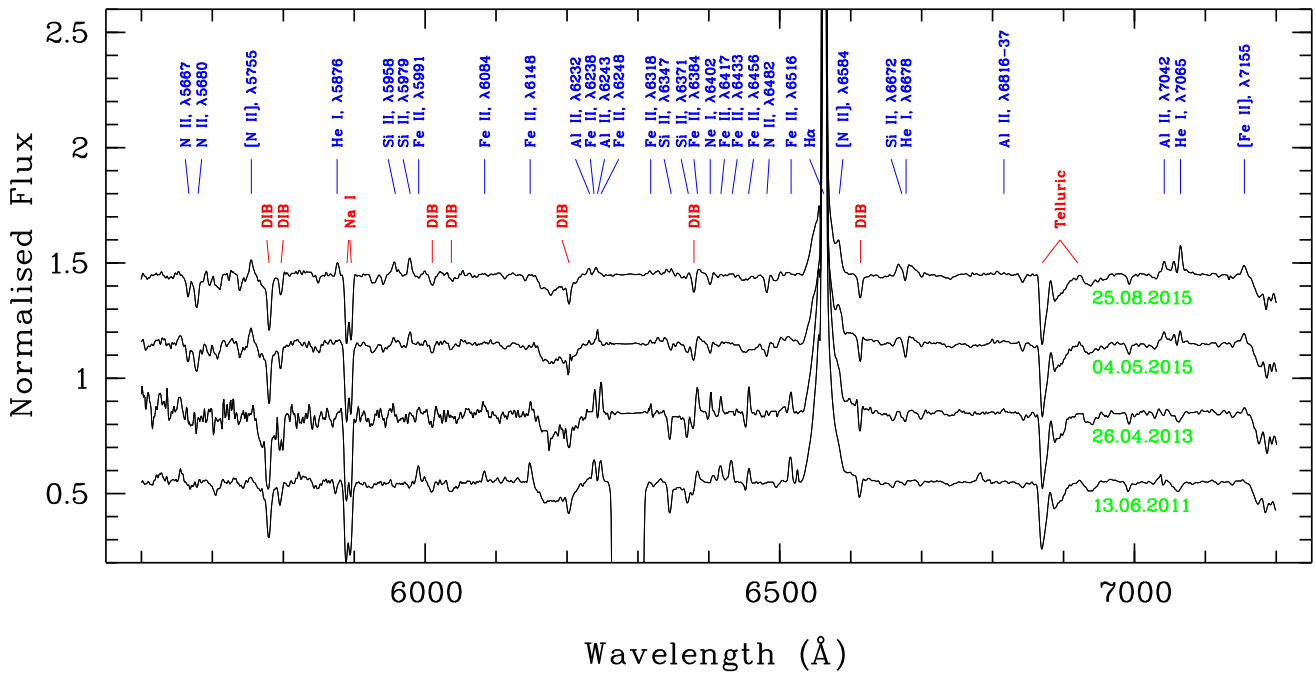}
\end{center}
\caption{Evolution of the (normalized) long-slit spectrum of MN48
in 2011--2015. Principal lines and most prominent DIBs are
indicated.} \label{fig:spec}
\end{figure*}

The nebula around MN48 was discovered in Gvaramadze et al. (2010b;
see their fig.\,1i) using the data of the {\it Spitzer}/MIPSGAL
survey (Carey et al. 2009). In the Set of Identifications,
Measurements and Bibliography for Astronomical Data (SIMBAD) data
base\footnote{http://simbad.harvard.edu/simbad/}, this nebula is
named [GKF2010]\,MN48. Like the majority of other mid-infrared
nebulae discovered with {\it Spitzer} and {\it WISE}, the nebula
around MN48 emerges in all its glory only in the {\it Spitzer}
24\,$\mu$m and {\it WISE} 22\,$\mu$m images\footnote{{\it Spitzer}
spectroscopy of circumstellar nebulae around evolved massive stars
showed that their emission is dominated by the dust continuum
emission (Flagey et al. 2011; Nowak et al. 2014). Detection of the
nebula around MN48 at longer wavelengths (e.g. in the {\it
Herschel} data or with the APEX/LABOCA) could, therefore, be used
to estimate its mass and to constrain the mass-loss history of the
star itself.}. At 24\,$\mu$m it appears (see the upper, left-hand
panel of Fig.\,\ref{fig:neb}) as a nearly circular (slightly
elongated southwest to northeast) shell of radius of $\approx$1.5
arcmin\footnote{The angular size of the nebula around MN48 is the
largest among 115 nebulae discovered in Gvaramadze et al.
(2010b).}, with a rather ragged rim. There are also some hints of
12 and 8\,$\mu$m emission (more obvious at 12\,$\mu$m) probably
associated with the western (more bright) edge of the nebula (see
the upper, middle and right-hand panels of Fig.\,\ref{fig:neb}).
The enhanced brightness of this part of the nebula might be caused
by the interaction of the nebula with the \hii region
IRAS\,16455$-$4531, which is located at $\approx$5 arcmin to the
west (the possible association between MN48 and IRAS\,16455$-$4531
is discussed in Section\,\ref{sec:hii}).

MN48 is prominent at 24\,$\mu$m, in all IRAC and {\it WISE}
wavebands, and in all ($J,H,K_{\rm s}$) Two-Micron All Sky Survey
(2MASS) images (Skrutskie et al., 2006). The 2MASS coordinates of
this star are: $\alpha_{2000}$=$16^{\rm h} 49^{\rm m} 37\fs70$,
$\delta_{J2000}$ =$-45\degr 35\arcmin 59\farcs3$ and
$l$=$340\fdg0298$, $b$=$-$$0\fdg5839$. More importantly, MN48 is
also visible in the optical wavebands [see the lower, right-hand
panel of Fig.\,\ref{fig:neb} for the Digitized Sky Survey II
(DSS-II) red band (McLean et al. 2000) image], which makes this
star a good target for our observing programmes with the Southern
African Large Telescope (SALT).

Wachter et al. (2010, 2011) classified MN48 (or star \#20 in their
designation) as a Be star by means of infrared spectroscopy. The
$H$ and $K$ band spectra of this star were presented in fig.\,6 of
Wachter et al. (2010).

\section{Observations of MN48}
\label{sec:obs}

\subsection{Long-slit SALT spectroscopy}
\label{sec:SALT}

\begin{table}
\caption{Journal of the long-slit observations.} \label{tab:Obs}
\begin{tabular}{llcc} \hline
Date & Exposure & Slit/Seeing & JD \\
& (min) & (arcsec) & (d) \\
 \hline
 2011 June 13        & 1$\times$10             &  1.00/1.8    &  2455725  \\
 2013 April 26       & 1$\times$10             &  1.25/1.5    &  2456408  \\
 2015 May 4          &
 1$\times$10  &  1.25/0.9    &  2457146  \\
 2015 August 25      & 
 1$\times$15  &  1.25/1.4    &  2457259  \\
 \hline
\end{tabular}
\end{table}

\begin{table}
\centering{ \caption{EWs, FWHMs and RVs of some lines in the
spectra of MN48.} \label{tab:inten}
\begin{tabular}{lrcr}
\hline \rule{0pt}{10pt}
& \MC{1}{c}{EW($\lambda$)} & FWHM($\lambda$) & \MC{1}{c}{RV} \\
$\lambda_{0}$(\AA) Ion & \MC{1}{c}{(\AA)} & \MC{1}{c}{(\AA)} & \MC{1}{c}{($\kms$)} \\
\hline  \rule{0pt}{10pt} & \MC{2}{c}{2011 June 13} \\
6347\ Si\ {\sc ii}\     &$-$0.73$\pm$0.05 &  6.05$\pm$0.33 &  $-$95$\pm$19 \\
6563\ H$\alpha$\        &  28.58$\pm$0.90 &  6.85$\pm$0.25 &  $-$21$\pm$18 \\
\\
\rule{0pt}{10pt} & \MC{2}{c}{2013 April 26} \\
6347\ Si\ {\sc ii}\     &$-$0.48$\pm$0.06 &  5.04$\pm$0.19 & $-$100$\pm$15 \\
6563\ H$\alpha$\        &  47.05$\pm$0.88 &  5.44$\pm$0.12 &      8$\pm$15 \\
7155\ [Fe\ {\sc ii}]\   &   0.08$\pm$0.05 &  4.12$\pm$0.54 &      7$\pm$16 \\
\\
\rule{0pt}{10pt} & \MC{2}{c}{2015 May 4} \\
4861\ H$\beta$\         &   2.91$\pm$0.19 &  6.98$\pm$0.54 &     75$\pm$15 \\
5755\ [N\ {\sc ii}]\    &   0.30$\pm$0.05 &  5.33$\pm$0.40 &      7$\pm$10 \\
6347\ Si\ {\sc ii}\     &$-$0.10$\pm$0.04 &  3.95$\pm$0.72 &  $-$27$\pm$12 \\
6563\ H$\alpha$\        &  22.21$\pm$0.55 &  5.19$\pm$0.15 &     24$\pm$15 \\
7065\ He\ {\sc i}\      &   0.26$\pm$0.05 &  4.18$\pm$0.24 &  $-$19$\pm$18 \\
7155\ [Fe\ {\sc ii}]\   &   0.20$\pm$0.05 &  6.75$\pm$0.34 &  $-$33$\pm$15 \\
\\
\rule{0pt}{10pt} & \MC{2}{c}{2015 August 25} \\
4861\ H$\beta$\          &  3.93$\pm$0.29 &  4.88$\pm$0.41 &  $-$32$\pm$12 \\
5755\ [N\ {\sc ii}]\     &  0.29$\pm$0.02 &  4.87$\pm$0.28 &   $-$9$\pm$8  \\
5876\ He\ {\sc i}\       &  0.23$\pm$0.02 &  4.12$\pm$0.30 &     20$\pm$8  \\
6347\ Si\ {\sc ii}\      &  0.02$\pm$0.01 &  4.69$\pm$0.37 &  $-$29$\pm$7  \\
6563\ H$\alpha$\         & 20.84$\pm$0.47 &  5.41$\pm$0.14 &      0$\pm$5  \\
7065\ He\ {\sc i}\       &  0.61$\pm$0.03 &  5.48$\pm$0.24 &  $-$32$\pm$6  \\
7155\ [Fe\ {\sc ii}]\    &  0.23$\pm$0.02 &  5.82$\pm$0.44 &  $-$54$\pm$9  \\
\\
\rule{0pt}{10pt} & \MC{2}{c}{2015 August 26 (\'echelle)} \\
6563\ H$\alpha$\        &  20.21$\pm$0.12 &  1.57$\pm$0.02 & $-$11.4$\pm$0.4   \\
6584\ [N\ {\sc ii}]\    &  0.43$\pm$0.02  &  5.29$\pm$0.17 & $-$28.2$\pm$3.3   \\
7065\ He\ {\sc i}\      &   0.60$\pm$0.01 &  2.17$\pm$0.06 & $-$38.2$\pm$1.1   \\
7772\ O\ {\sc i}\       &$-$0.64$\pm$0.01  &  2.69$\pm$0.03 &$-$43.4$\pm$0.3   \\
7774\ O\ {\sc i}\       &$-$0.41$\pm$0.01  &  1.61$\pm$0.04 &$-$29.3$\pm$0.3   \\
7775\ O\ {\sc i}\       &$-$0.05$\pm$0.01  &  0.72$\pm$0.05 &$-$28.4$\pm$0.3   \\
7877\ Mg\ {\sc ii}\     &   0.50$\pm$0.01 &  1.57$\pm$0.03 & $-$45.5$\pm$0.5   \\
7896\ Mg\ {\sc ii}\     &   0.61$\pm$0.01 &  1.74$\pm$0.02 & $-$43.7$\pm$0.4   \\
\hline \MC{4}{p{8cm}}{{\it Note}: The negative EWs correspond to
absorption lines.}
\end{tabular}
 }
\end{table}

MN48 was observed with the SALT (Buckley, Swart \& Meiring 2006;
O'Donoghue et al. 2006) on four occasions in 2011--2015 (see
Table\,\ref{tab:Obs} for the log of the observations). The spectra
were taken with the Robert Stobie Spectrograph (RSS; Burgh et al.
2003; Kobulnicky et al. 2003) in the long-slit mode. The PG900
grating was used for all observations, and in all cases the
spectral range of $4200-7300$~\AA\ was covered with a final
reciprocal dispersion of $0.97$~\AA\ pixel$^{-1}$. The spectral
resolution full width at half-maximum (FWHM) was of
$4.40\pm$0.15~\AA. Because the spectra were taken during bright
time, their blue parts (shortward of $\approx$5600\, \AA) are very
noisy and do not contain meaningful information. A Xe lamp arc
spectrum was taken immediately after all science frames.
Spectrophotometric standard stars were observed during twilight
time for relative flux calibration. Absolute flux calibration is
not feasible with SALT because the unfilled entrance pupil of the
telescope moves during the observations.

Primary reduction of the data was done in the standard way with
the SALT science pipeline (Crawford et al. 2010). Following
long-slit data reduction was carried out in the way described in
Kniazev et al. (2008). The resulting normalized spectra are
presented in Fig.\,\ref{fig:spec}. The principal lines and most
prominent diffuse interstellar bands (DIBs) are indicated. All
wavelengths are given in air. Equivalent widths (EWs), FWHMs and
heliocentric radial velocities (RVs) of some lines in the spectra
(measured applying the {\sc midas} programs; see Kniazev et al.
2004 for details) are given in Table\,\ref{tab:inten}.

\begin{figure}
\begin{center}
\includegraphics[angle=-90,width=1.0\columnwidth,clip=]{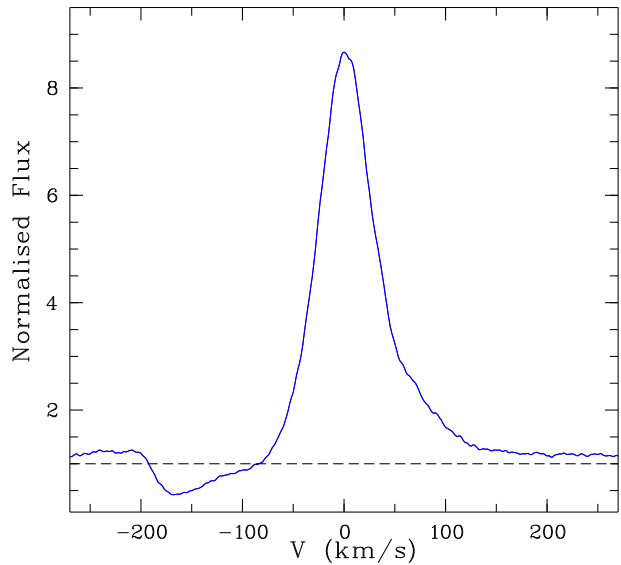}
\end{center}
\caption{The H$\alpha$ line profile in the \'echelle spectrum. The
blue edge of the absorption component corresponds to the terminal
wind velocity of $\approx$$200 \, \kms$.} \label{fig:Ha}
\end{figure}

\subsection{\'Echelle spectroscopy with the SALT} \label{sec:HRS}

\begin{figure*}
\begin{center}
\includegraphics[angle=270,width=0.4\textwidth,clip=]{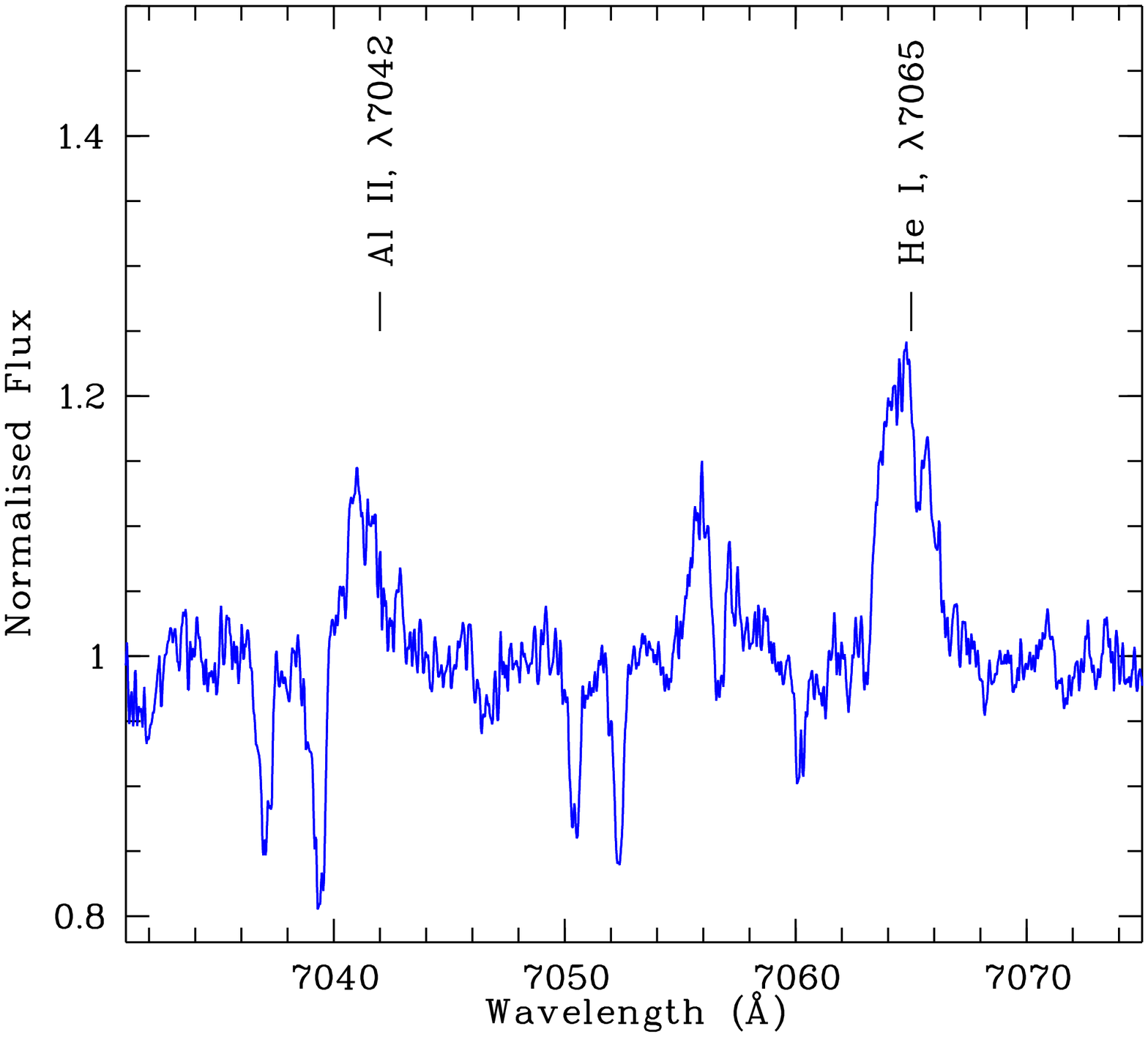}
\includegraphics[angle=270,width=0.4\textwidth,clip=]{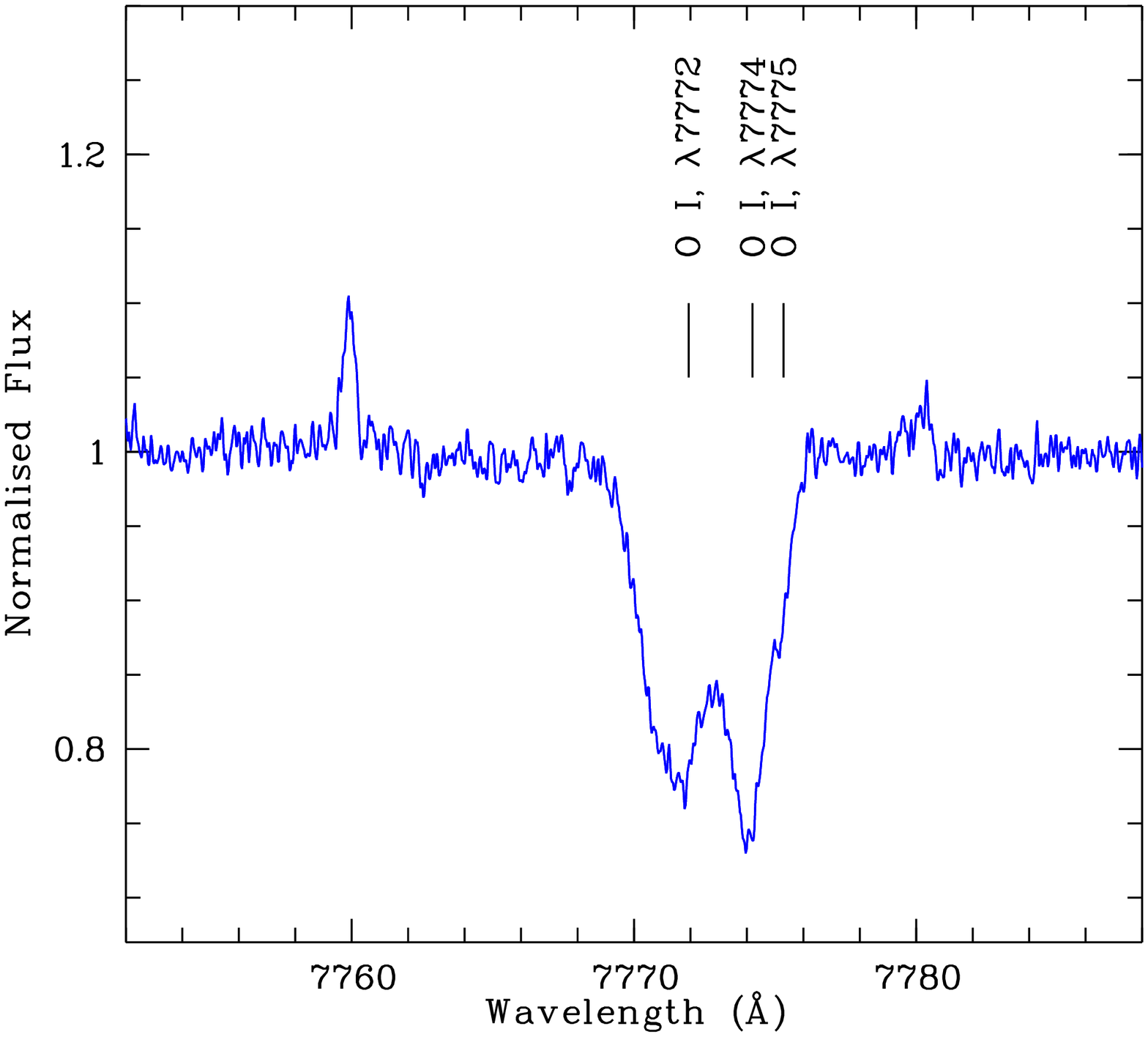}
\includegraphics[angle=270,width=0.4\textwidth,clip=]{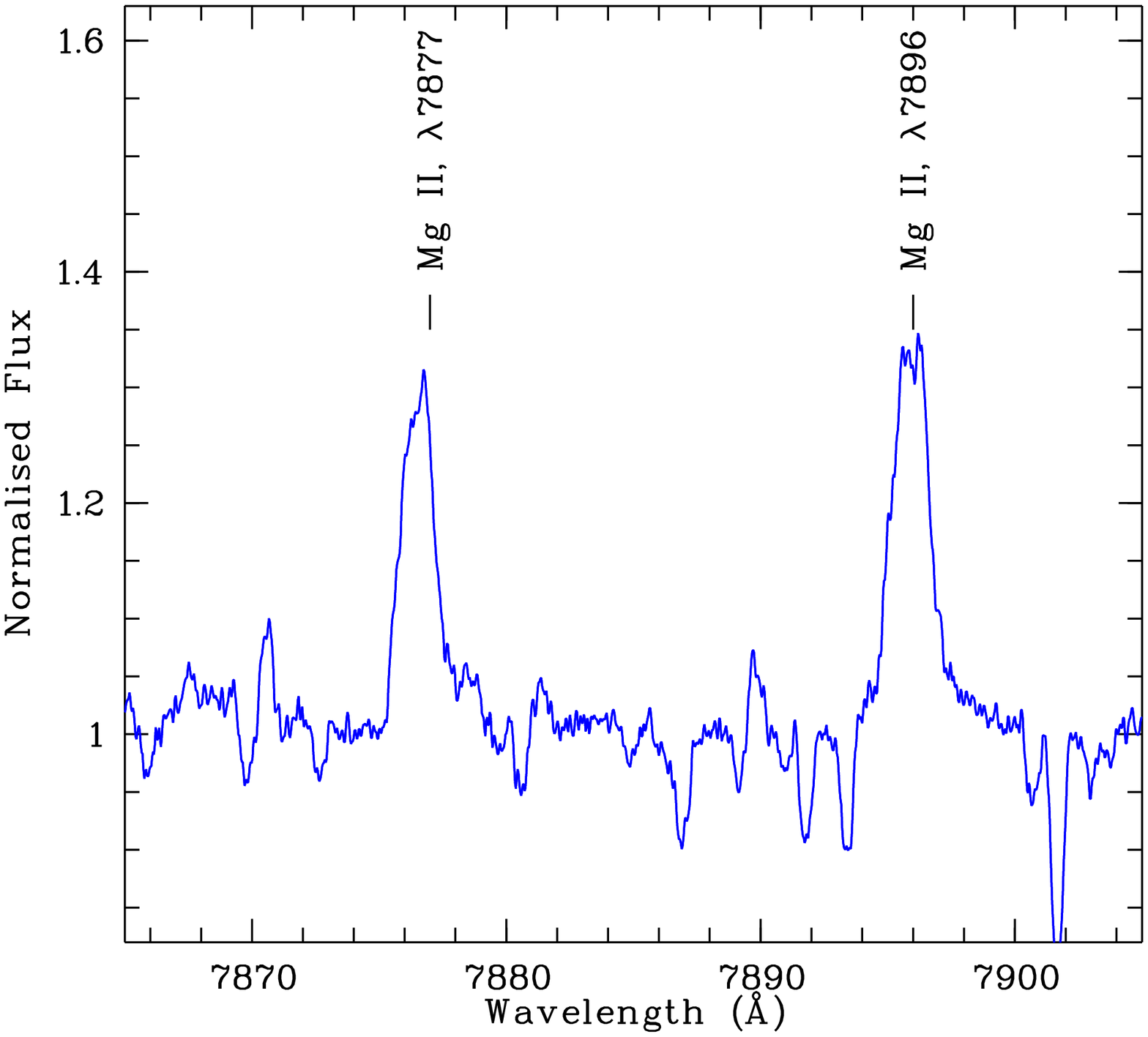}
\includegraphics[angle=270,width=0.4\textwidth,clip=]{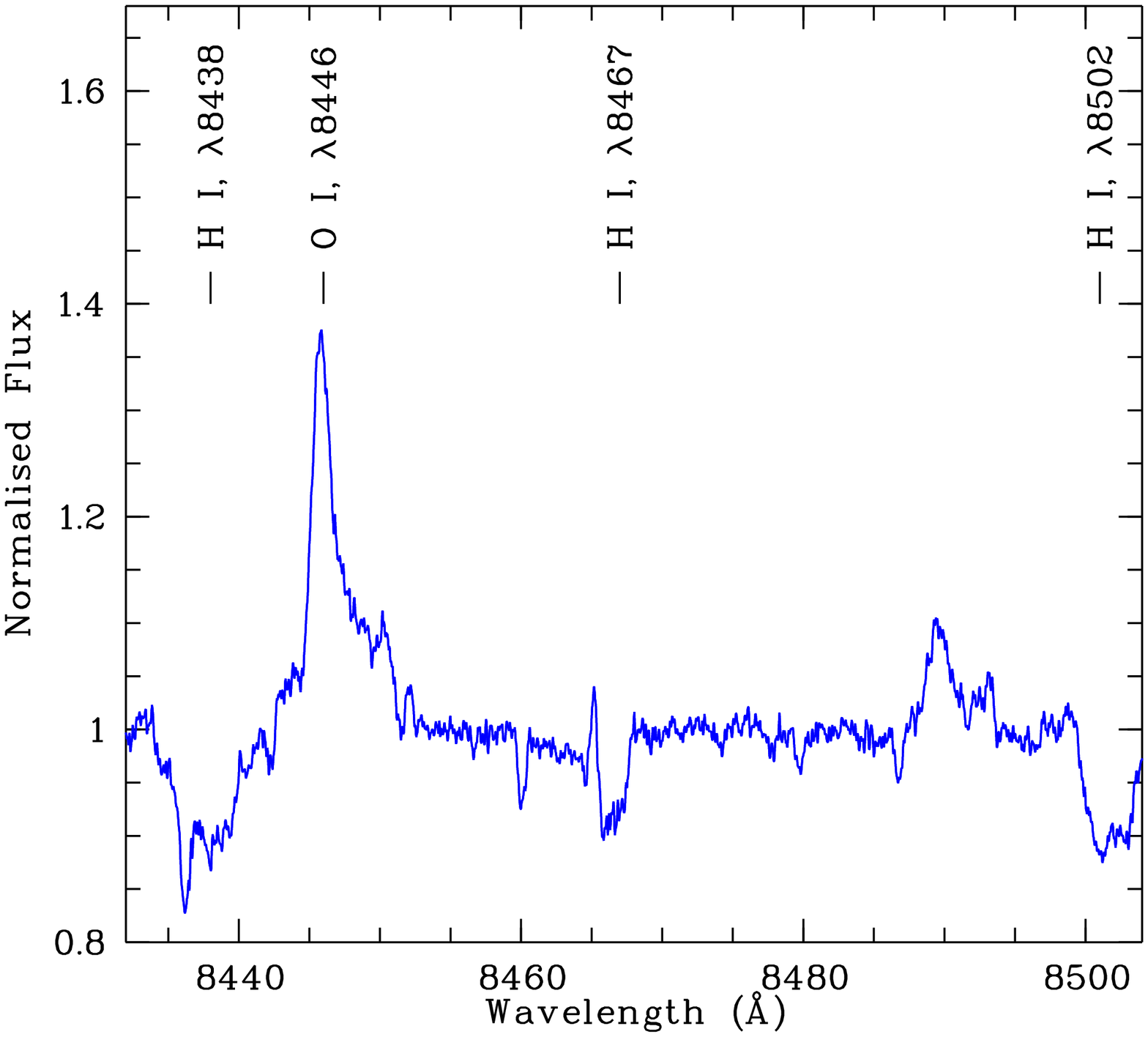}
\end{center}
\caption{Portions of the \'echelle spectrum of MN48 (obtained on
2015 August 26). Narrow unlabeled absorption lines (e.g. at 7037,
7040, 7049 and 7051 \AA) are telluric in origin.} \label{fig:ech}
\end{figure*}

MN48 was also observed with the SALT High Resolution Spectrograph
(HRS; Barnes et al. 2008; Bramall et al. 2010, 2012; Crause et al.
2014) on 2015 August 26 with a single exposure of 1800 s and
seeing of about 3 arcsec. The HRS is a dual beam, fibre-fed
\'echelle spectrograph. It was used in the low resolution mode
(R$\sim$14000 and 2.23 arcsec diameter for both the object and sky
fibers) to provide a spectrum in the blue and red arms over the
total spectral range of $\approx$3700--8900~\AA. The both blue and
red arms CCD were read out by a single amplifier with a 1$\times$1
binning. Three arc spectra of ThAr+Ar lamps and three spectral
flats were obtained in this mode during a weekly set of HRS
calibrations. MN48 is a highly reddened object and for this reason
the blue arm spectrum is unusable. The red arm spectrum has a
signal-to-noise ratio per pixel of $\geq$25 in the middle of
spectral orders starting from $\sim$6300~\AA \, and could be used
for the brightest longward lines. Because of problems with the CCD
in the red arm science frame, however, a part of the spectrum
beyond $\approx$8500~\AA\ was unusable for analysis.

Primary reduction of the HRS data, including overscan correction,
bias subtractions and gain correction, was done with the SALT
science pipeline (Crawford et al. 2010). Spectroscopic reduction
of the red arm data was carried out using standard {\sc midas}
\texttt{feros} (Stahl, Kaufer \& Tubbesing 1999) and
\texttt{echelle} (Ballester 1992) packages. We have performed the
following steps: (1) positions for 33 spectral orders (from 53
till 85) for both fibers were found using spectral flats frames;
(2) the 2D background was determined and subtracted from all
frames; (3) the straightened \'echelle spectrum was extracted for
both fibers from all types of frames (flats, arcs and object)
using the standard mode with cosmic masking and the optimum
extraction algorithms; (4) for the object and sky fibers the blaze
function was removed from the science frame through division to
extracted spectrum of spectral flat; (5) the procedure found
$\sim$1100 emission lines in the extracted arc spectrum of which
$\sim$450 lines were finally automatically identified with
requested level of tolerance to build a 2D dispersion curve with
the final mean rms of 0.006~\AA (this step was done independently
for the object and sky fibers); (6) all extracted orders were
rebinned into linear wavelength steps and the wavelength
calibrated sky fiber orders were subtracted from the object fiber
orders; (7) all orders for the object fiber were merged into a 1D
file. The total spectral range of 5400--8500~\AA\ was covered with
a final reciprocal dispersion of $0.04$~\AA\ pixel$^{-1}$. The
spectral resolution FWHM (measured using wavelength calibrated arc
spectrum) changes from $\approx$0.34~\AA\ in the blue part of the
spectrum (R$\sim$16500) to $\approx$0.57~\AA\ in the red part
(R$\sim$15500).

EWs, FWHMs and RVs of some lines in the HRS spectrum are given in
Table\,\ref{tab:inten}. We estimated the accuracy of the continuum
normalization to be better than 5 per cent. The comparison of the
EWs of the H$\alpha$ line in the RSS and HRS spectra obtained
during two sequential nights supports this estimate. Several
portions of the \'echelle spectrum (not corrected for barycentric
motion) are presented in Figs\,\ref{fig:Ha} and \ref{fig:ech}.
Fig.\,\ref{fig:Ha} shows that the H$\alpha$ line has a clear
P\,Cygni profile. Note that the line profiles of the Paschen lines
seen in the lower right panel of Fig.\,\ref{fig:ech} are quite
different from those observed in the spectra of the classical LBVs
like P\,Cygni (Stahl et al. 1993) and AG\,Car (Groh et al. 2009).
Better observational data are required to check whether this
distinction is real or is due to instrumental effects.

\subsection{Photometry}
\label{sec:phot}

\begin{table*}
  \caption{Archival and contemporary photometry of MN48.}
  \label{tab:phot}
    \begin{tabular}{lccccccl}
      \hline
      Date                  & $B$            & $V$            & $I_{\rm c}$    & $V-I_{\rm c}$ & $J$           & $K_{\rm}$     & JD \\
      \hline
      1976 April 30$^a$     & 18.6$\pm$0.3   & --             & --             & --            & --            & --            & 244\,2899 \\
      1980 July 22$^a$      & --             & --             & 11.20$\pm$0.20 & --            & --            & --            & 244\,4442 \\
      1998 July 28$^b$      & --             & --             & 10.82$\pm$0.02 & --            & 7.87$\pm$0.10 & 4.54$\pm$0.12 & 245\,1023 \\
      1999 May 23$^c$       & --             & --             & --             & --            & 7.24$\pm$0.02 & 5.42$\pm$0.02 & 245\,1322 \\
      1999 June 25$^b$      & --             & --             & 11.07$\pm$0.03 & --            & 7.11$\pm$0.10 & 4.86$\pm$0.09 & 245\,1355 \\
      2009 April 18$^d$     & --             & 16.05$\pm$0.05 & 10.74$\pm$0.02 & 5.31$\pm$0.05 & --            & --            & 245\,4940 \\
      2011 May 10$^e$       & 18.18$\pm$0.10 & 15.19$\pm$0.03 &  9.73$\pm$0.01 & 5.46$\pm$0.03 & --            & --            & 245\,5692 \\
      2012 May 6$^e$        & --             & 15.26$\pm$0.01 &  9.93$\pm$0.01 & 5.33$\pm$0.02 & --            & --            & 245\,6054 \\
      2013 January 24$^e$   & --             & 15.61$\pm$0.02 & 10.29$\pm$0.01 & 5.32$\pm$0.02 & --            & --            & 245\,6316 \\
      2014 April 25$^e$     & 19.27$\pm$0.20 & 15.92$\pm$0.05 & 10.69$\pm$0.04 & 5.23$\pm$0.06 & --            & --            & 245\,6772 \\
      2015 May 4$^f$        & --             & --             & 11.26$\pm$0.08 & --            & --            & --            & 245\,7146 \\
      2015 May 19$^g$       & --             & 16.33$\pm$0.03 & 11.19$\pm$0.04 & 5.14$\pm$0.05 & --            & --            & 245\,7161 \\
      2015 August 25$^f$    & --             & --             & 11.33$\pm$0.06 & --            & --            & --            & 245\,7259 \\
      2016 February 2$^h$   & --             & 16.24$\pm$0.03 & 11.10$\pm$0.03 & 5.14$\pm$0.04 & --            & --            & 245\,7420 \\
      \hline
  \MC{8}{l}{$^{a}$USNO B-1; $^b$DENIS; $^{c}$2MASS; $^{d}$40-cm telescope; $^e$76-cm telescope; $^f$SALT; $^g$1-m telescope; $^h$LCOGT.}
    \end{tabular}
    \end{table*}

To search for photometric variability of MN48, we occasionally
obtained its CCD photometry with the 40-cm telescope of the
Observatorio Cerro Armazones (OCA) of the North Catholic
University (Chile) and the 76-cm telescope of the South African
Astronomical Observatory during our observing runs in 2009--2014.
We used an SBIG ST-10XME CCD camera equipped with $BVI_{\rm c}$
filters of the Kron-Cousins system (for details see Berdnikov et
al. 2012) to build the system of secondary standards in the field
of MN48. With these standards, we derived $V$ and $I_{\rm c}$
magnitudes of MN48 using images obtained with the SAAO 1-m
telescope on 2015 May 19, an $I_{\rm c}$ magnitude using
acquisition images obtained with the SALT on 2015 May 4 and August
25, and $V$ and $I_{\rm c}$ magnitudes from images obtained with
the Las Cumbres Observatory Global Telescope (LCOGT) network
(Brown et al. 2013) on 2016 February 2. Using the same standards,
we recalibrated the $I$ and $B$ magnitudes from the USNO\,B-1
catalogue (Monet et al. 2003). The results are presented in
Table\,\ref{tab:phot}. To this table we also added the $J$ and
$K_{\rm s}$ magnitudes from 2MASS (Cutri et al. 2003) and
two-epoch $I, J$ and $K$ photometry from the Deep Near Infrared
Survey of the Southern Sky (DENIS; Epchtein et al. 1999).

\begin{figure}
\begin{center}
\includegraphics[angle=-90,width=1.0\columnwidth,clip=]{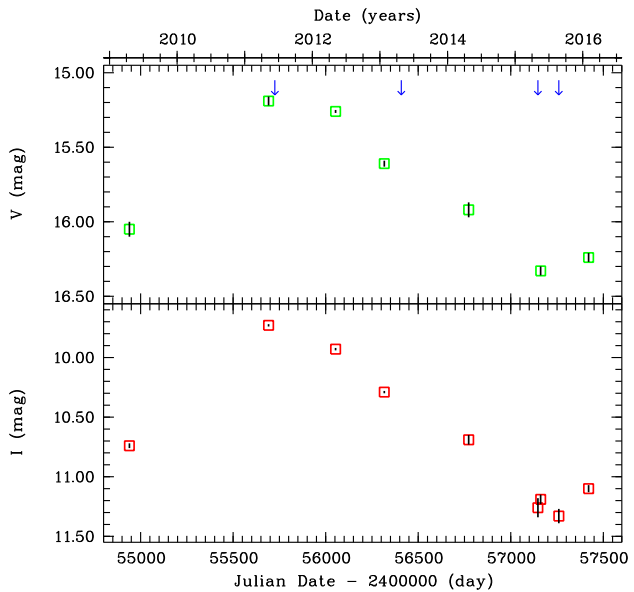}
\end{center}
\caption{Light curves of MN48 in the $V$ and $I$ bands in
2009--2016. 1$\sigma$ error bars are indicated, but in most cases
they are within the size of the data points (boxes). The arrows
mark the dates of the SALT spectra.} \label{fig:phot}
\end{figure}

The $V$ and $I$ band light curves of MN48, based on our
observations in 2009--2016, are presented in Fig.\,\ref{fig:phot}.
The arrows indicate times when the spectra were obtained with the
SALT. For each data point (square) we give 1$\sigma$ error bars,
which in most cases are less than the size of the squares
themselves and therefore they cannot be discerned.

\begin{figure}
\begin{center}
\includegraphics[angle=-90,width=1.0\columnwidth,clip=]{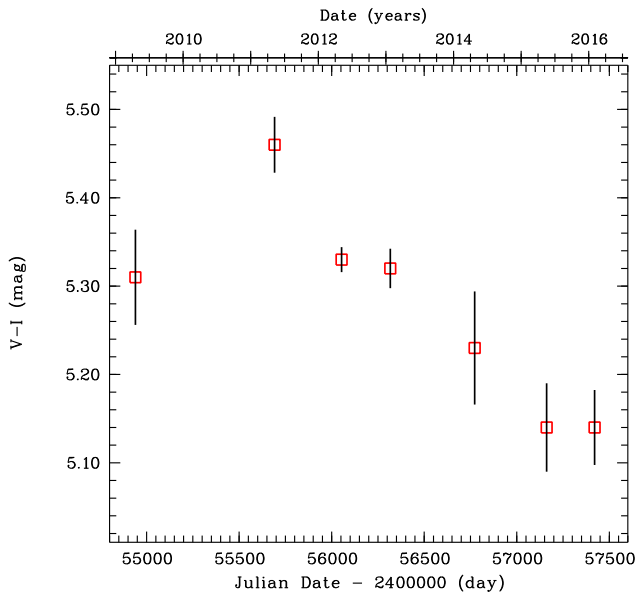}
\end{center}
\caption{Evolution of the $V$$-$$I_{\rm c}$ colour of MN48 with
time.} \label{fig:col}
\end{figure}

\section{Discussion}
\label{sec:dis}

\subsection{MN48: a new Galactic bona fide LBV}
\label{sec:lbv}

Inspection of Table\,\ref{tab:phot} and Fig.\,\ref{fig:phot} shows
that MN48 has brightened in the $V$ and $I_{\rm c}$ bands,
respectively, by $\approx$0.9 and 1 mag during the first two years
of our photometric monitoring campaign and reached the maximum
light by the time of our first spectroscopic observation in 2011.
Since then, MN48 started to fade and its brightness in the $V$ and
$I_{\rm c}$ bands has dropped, respectively, by $\approx$1.1 and
1.6 mag during the next four years. The last data points (obtained
on 2016 February 2) suggest that MN48 has started to brighten
again.

The photometric variability of MN48 could also be inferred from
comparison of the $I_{\rm c}$ magnitudes based on our observations
with those given in the USNO\,B-1 and DENIS catalogues, as well as
from comparison of the two epoch DENIS $J$ and $K$ photometry with
the 2MASS photometry (see Table\,\ref{tab:phot}). Interestingly,
while the $J$ band brightness of MN48 increased by $\approx$0.8
mag during $\approx$11 months in 1998--1999, the $K$ band
brightness instead faded by $\approx$0.9 mag during the first 10
months and then increased by $\approx$0.6 mag during the next
month. Also, comparison of the IRAC 4.5\,$\mu$m magnitude of MN48
of 5.02$\pm$0.06 (Spitzer Science Center 2009) with the {\it WISE}
4.6\,$\mu$m magnitude of 3.63$\pm$0.14 (Cutri et al. 2014)
indicates that the star became brighter by a factor of about 3 on
a time scale of four years.

From Table\,\ref{tab:phot} also follows that the $V-I_{\rm c}$
colour of MN48 has increased (i.e. the star became redder) with
the brightness increase and then started to decrease with the
brightness decline (i.e. the star becomes bluer). These changes in
the colour (see Fig.\,\ref{fig:col}) suggest that the effective
temperature, $T_{\rm eff}$, of MN48 has decreased during the first
two years of our photometric observations and after reaching the
minimum value by 2011 June has started to increase (e.g. van
Genderen 1982). This behaviour is typical of LBVs experiencing
S\,Dor-like outbursts and it should be accompanied by noticeable
changes in the spectral appearance of the star (e.g. Stahl et al.
2001).

Indeed, inspection of the RSS spectra presented in
Fig.\,\ref{fig:spec} confirms that MN48 became hotter during the
last four years, which is manifested in the almost complete
disappearance of the Fe\,{\sc ii} lines in the 2015's spectra (cf.
Stahl et al. 2001) and significant changes in the temperature
sensitive lines like N\,{\sc ii} $\lambda\lambda$5667, 5680, 6482,
Si\,{\sc ii} $\lambda\lambda$6347, 6371, and Ne\,{\sc ii}
$\lambda$6402 (e.g. Walborn 1980; Lennon, Dufton \& Fitzsimmons
1993). Particularly, the Si\,{\sc ii} $\lambda\lambda$6347, 6371
doublet was very prominent in 2011 and 2013 and is hardly visible
in 2015, which implies that $T_{\rm eff}$ of MN48 has increased by
several thousand degrees during the last two years (e.g. Lennon et
al. 1993).

\begin{figure}
\begin{center}
\includegraphics[angle=-90,width=1.0\columnwidth,clip=]{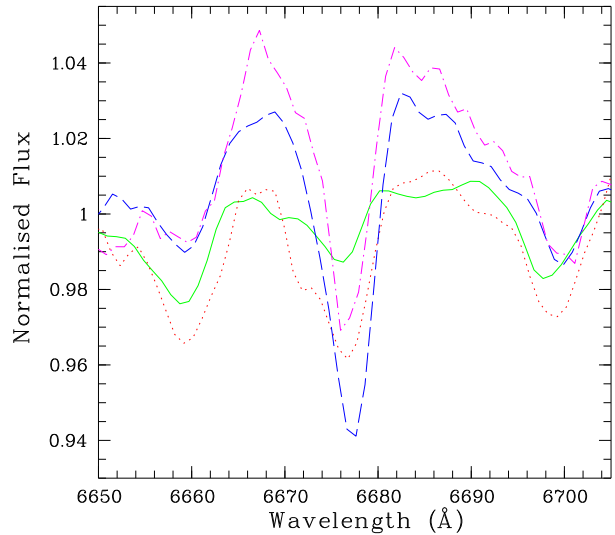}
\end{center}
\caption{Evolution of the He\,{\sc i} $\lambda$6678 line profile
with time: 2011 June 13 (solid green line), 2013 April 26 (dotted
red line), 2015 May 4 (dashed blue line) and 2015 August 25
(dash-dotted magenta line).} \label{fig:6678}
\end{figure}

The brightness decline of MN48 is accompanied by appearance of
forbidden lines of [N\,{\sc ii}] $\lambda\lambda$5755, 6584 and
[Fe\,{\sc ii}] $\lambda$7155, which became prominent in the 2015's
spectra. The spectra also show significant changes in the profiles
of the He\,{\sc i} $\lambda\lambda$6676, 7065 lines. These lines
were weakly in absorption in the 2011's spectrum, but increased
their strength in 2013, then showed up prominent blue and red
emission wings in the first 2015's spectrum, which became even
more prominent in the second 2015's spectrum (see
Fig.\,\ref{fig:6678}).

The \'echelle spectrum of MN48 shows that the [Fe\,{\sc ii}]
$\lambda$7155 line has a flat-topped profile. This indicates that
the line is formed in a region of constant expansion velocity and
therefore its width is a measure of the terminal wind velocity,
$v_\infty$ (Stahl et al. 1991). Using this line, we derived
$v_\infty$=152$\pm$5$ \, \kms$. This figure is comparable to the
wind velocity estimate of $\approx$$200 \, \kms$ based on the blue
edge of the absorption component of the H$\alpha$ line (see
Fig.\,\ref{fig:Ha}).

The \'echelle spectrum also shows the prominent O\,{\sc i}
$\lambda$7772--5\,\AA \, triplet in absorption
(Fig.\,\ref{fig:ech}), which is a good luminosity indicator
(Merrill 1934). The EW of this triplet of 1.10$\pm$0.02 \AA \,
(see Table\,\ref{tab:inten}) implies that MN48 is a B-type
supergiant (e.g. Keenan \& Hynek 1950). But unlike the normal blue
supergiants of the same temperature, the wind of MN48 is slower
(cf. Crowther 1997), which could be considered as an indication
that this star is close to the Eddington limit (Groh, Hillier \&
Damineli 2011; Humphreys et al. 2014).

The [Fe\,{\sc ii}] $\lambda$7155 line can also be used to estimate
the systemic velocity of MN48 (Stahl et al. 2001). Using the
\'echelle spectrum, we found a heliocentric RV of this line of
$\approx$$-37 \, \kms$. Interestingly, this RV is equal to the
mean RV of other lines in the \'echelle spectrum (listed in
Table\,\ref{tab:inten}), which are not affected by P\,Cygni
absorption. We adopt this figure as a systemic velocity of MN48.

To summarize, the detected changes in the brightness and spectrum
of MN48 imply that this star shows currently an S\,Dor activity
and therefore it is one more (18th) example of the Galactic bona
fide LBVs. A census of these objects is given in table\,2 of
Kniazev \& Gvaramadze (2016).

\subsection{Distance to and luminosity of MN48}
\label{sec:dist}

The majority of LBVs are located beyond the confines of known star
clusters and therefore they are likely runaway stars (Gvaramadze
et al. 2012b). Just the isolation from other massive stars allows
the LBVs to create coherent (observable) circumstellar nebulae,
while within the parent clusters formation of nebulae is precluded
because of the effect of stellar winds from neighbouring OB stars.
On the other hand, the isolated location of the majority of LBVs
makes their distances difficult to determine. Nonetheless, to get
some idea on the distance to a particular isolated LBV, one can
consider whether or not its luminosity would be reasonable if the
star is placed in one or another spiral arm along the line of
sight to it (cf. Gvaramadze et al. 2015b). Further constraint on
the distance to an isolated LBV could be obtained from detection
of its likely parent cluster.

To constrain the distance to (and hence the luminosity of) MN48,
we note that the sightline towards this star first enters (e.g.
Cordes \& Lazio 2002) the Carina-Sagittarius spiral arm (located
at a distance of $d$$\sim$2 kpc from the Sun), then (at $d$$\sim$3
kpc) intersects the Crux-Scutum arm, and then twice crosses the
Norma arm (at $d$$\sim$5 and 12 kpc). The distances at which the
sightline intersects the spiral arms correspond to distance
modulus (DM) values of $\approx$11.5, 12.4, 13.5 and 15.4 mag,
respectively. To estimate the absolute visual magnitude, $M_V$, of
MN48 for a given DM, one needs to know the visual extinction,
$A_V$, towards this star.

To estimate $A_V$, we match the dereddened spectral slope of MN48
with those of stars of similar $T_{\rm eff}$ from the Stellar
Spectral Flux Library by Pickles (1998). To estimate the
temperature of MN48, we note that the 2015's spectra of this star
are very similar to those of the classical LBV AG\,Car in 2003
January (see fig.\,3 in Groh et al. 2009), when its $T_{\rm eff}$
was about 14\,000 K. Adopting this $T_{\rm eff}$ for MN48 and
using the RSS spectra, we found the colour excess of
$E(B-V)$=3.9$\pm$0.1 mag\footnote{Note that this estimate only
slightly depends on the adopted $T_{\rm eff}$ (Gvaramadze et al.
2012a).} and $A_V$=12.1$\pm$0.3 mag (here we used the ratio of
total to selective extinction of $R_V=3.1$). With this $A_V$ and
$V$=15.7$\pm$0.5 mag (see Table\,\ref{tab:phot}), and using the
above values of DM, one finds, respectively,
$M_V$$\approx$$-$7.9$\pm$0.6, $-$8.8$\pm$0.6, $-$9.9$\pm$0.6 and
$-$11.8$\pm$0.6 mag. These absolute magnitudes translate to the
luminosities of $\log(L_{\rm bol}/\lsun)$$\approx$5.5$\pm$0.2,
5.9$\pm$0.2, 6.3$\pm$0.2 and 7.1$\pm$0.2, respectively; here we
assumed that the bolometric correction of MN48 is equal to that of
AG\,Car in 2003 January (see above), i.e. $\approx$$-$1.2 mag
(Groh et al. 2009). One can see that the luminosity of MN48 would
be unreasonably high if this star is located in the far segment of
the Norma arm ($d=12$ kpc) or further out. The other three
luminosities are reasonable and imply that MN48 is located either
on the S\,Dor instability strip or on the cool side of this strip
(Wolf 1989; Groh et al. 2009) on the Hertzsprung-Russell diagram.
Note that the high extinction towards MN48 argues against the
short distance ($d$$\sim$2 kpc) to this star, although this
distance cannot completely be ruled out because a significant
fraction of reddening of MN48 might be due to the dusty
circumstellar material (cf. Gvaramadze et al. 2015b). We conclude,
therefore, that it is likely that MN48 is located either in the
Crux-Scutum arm or in the near segment of the Norma arm. In both
cases, the luminosity of MN48 would exceed the Humphreys-Davidson
luminosity limit (Humphreys \& Davidson 1979) and the star would
be subject to unsteady eruptive events.

The distance to MN48 could also be estimated through the systemic
velocity of this star. This estimate, however, would be reasonable
only if MN48 is not a runaway star with a significant peculiar
radial velocity. Using RV of MN48 of $-$$37 \, \kms$ and assuming
the distance to the Galactic Centre of $R_0$=8.0 kpc and the
circular rotation speed of the Galaxy of $\Theta _0 =240 \, \kms$
(Reid et al. 2009), and the solar peculiar motion
$(U_{\odot},V_{\odot},W_{\odot})=(11.1,12.2,7.3) \, \kms$
(Sch\"onrich, Binney \& Dehnen 2010), one finds the near and far
kinematic distances to MN48 of $\approx$2.7 and 12.3 kpc. The
former figure implies that MN48 is located in the interarm region,
which would be possible only if this star is a high-velocity
runaway moving almost along our line of sight (in this case,
however, the kinematic distance estimate would be meaningless). At
the latter distance, as noted above, the luminosity of MN48 would
be unrealistically high. These considerations suggest that either
the adopted systemic velocity of MN48 is incorrect or the star has
a peculiar radial velocity, i.e. it is a runaway star (this
possibility is further discussed in Sections\,\ref{sec:hii} and
\ref{sec:wd1}).

\subsection{MN48 and the \hii region IRAS\,16455$-$4531}
\label{sec:hii}

\begin{figure}
\begin{center}
\includegraphics[width=8cm,angle=0]{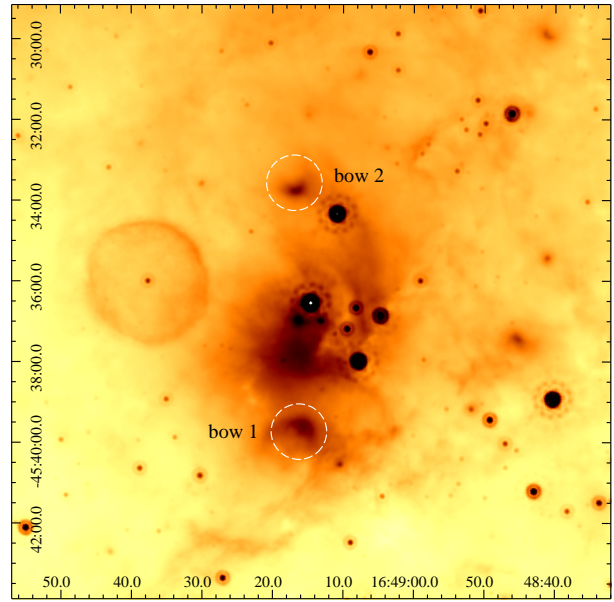}
\end{center}
\caption{{\it Spitzer} MIPS 24\,$\mu$m image of MN48 and the \hii
region IRAS\,16455$-$4531. The dashed circles mark two bow-shaped
structures faced towards the centre of IRAS\,16455$-$4531. The
coordinates are in units of RA (J2000) and Dec. (J2000) on the
horizontal and vertical scales, respectively. See text for
details.} \label{fig:iras}
\end{figure}

In Section\,\ref{sec:neb}, we mentioned that MN48 is located not
far from the \hii region IRAS\,16455$-$4531 and that the enhanced
brightness of the western edge of the nebula might be caused by
the interaction of the nebula with the \hii region. To discuss
this possibility, we present in Fig.\,\ref{fig:iras} the MIPS
24\,$\mu$m image of IRAS\,16455$-$4531 and its close environment.
On this image, besides MN48, one can see two bow-shaped structures
(hereafter, bow\,1 and bow\,2) pointed towards the centre of the
\hii region. We interpret these structures as bow shocks produced
because of interaction between a gas flow driven by a putative
star cluster deeply embedded in the \hii region and stellar winds
of two (massive) stars in the halo of the cluster (cf. Povich et
al. 2008). If our interpretation is correct, then the star cluster
embedded in IRAS\,16455$-$4531 might be the parent cluster to MN48
as well.

\begin{figure}
\begin{center}
\includegraphics[width=8cm,angle=0]{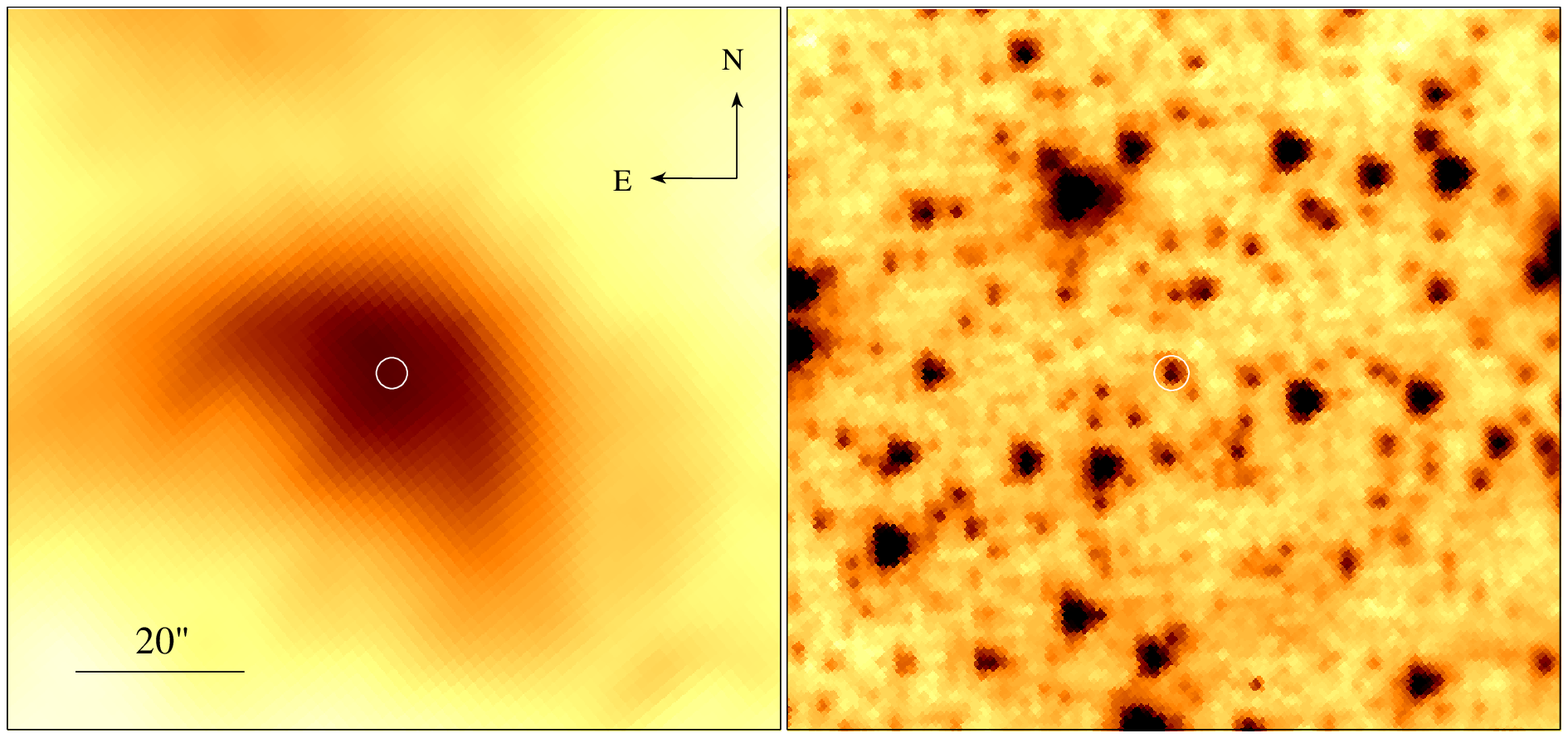}
\includegraphics[width=8cm,angle=0]{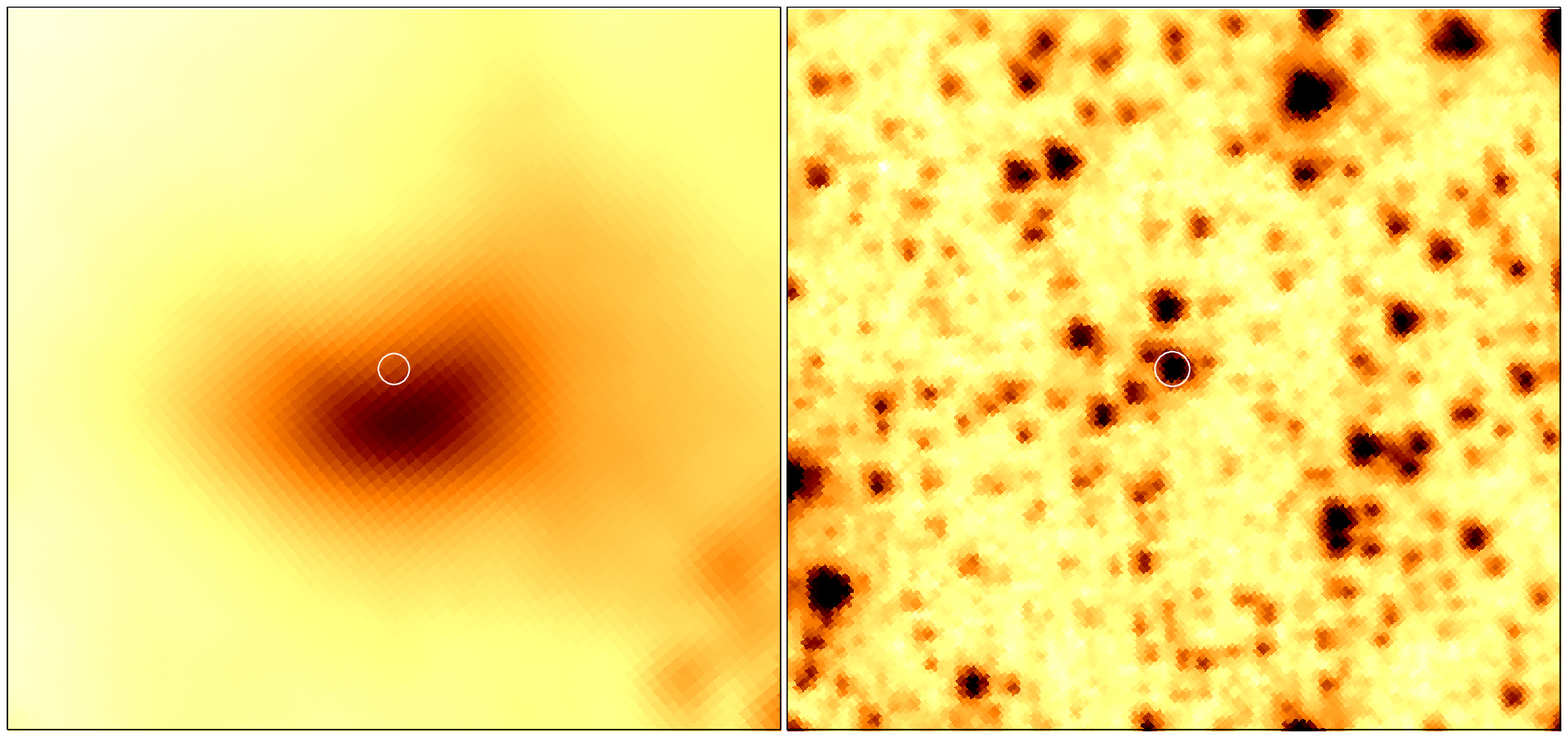}
\end{center}
\caption{Left panels: MIPS 24\,$\mu$m images of bow\,1 (upper
panel) and bow\,2. Right panels: IRAC 3.6\,$\mu$m images of the
candidate bow-shock-producing stars (marked by white circles):
star\,1 (upper panel) and star\,2. The orientation and the scale
of the images are the same.} \label{fig:bow}
\end{figure}

\begin{table*}
\caption{Details of two candidate bow-shock-producing stars around
IRAS\,16455$-$4531. See text for details.} \label{tab:VVV}
\begin{tabular}{cccccccc} \hline
star & RA(J2000) & Dec.(J2000) & $J$ (mag) & $K$ (mag) & $A_K$ (mag) & $M_K$ (mag) & Spectral type \\
\hline
1 & $16^{\rm h} 49^{\rm m} 15\fs65$ & $-$$45\degr 39\arcmin 35\farcs9$ & 17.103$\pm$0.025 & 13.299$\pm$0.008 & 2.65 & $-$2.84 & B1\,V \\
2 & $16^{\rm h} 49^{\rm m} 16\fs81$ & $-$$45\degr 33\arcmin 39\farcs1$ & 13.478$\pm$0.002 & 11.380$\pm$0.002 & 1.52 & $-$3.63 & O8\,V \\
\hline
\end{tabular}
\end{table*}

In Fig.\,\ref{fig:bow} we show stars possibly associated with
bow\,1 and bow\,2. These stars (hereafter, star\,1 and star\,2)
are highly extincted and visible only in the infrared. In
Table\,\ref{tab:VVV}, we give their coordinates and the $J$ and
$K$ band photometry from the VISTA Variables in the V\'{i}a
L\'{a}ctea (VVV) survey Catalogue Data Release 1 (Saito et al.
2012). To produce the bow shocks, stars\,1 and 2 should possess
strong winds, i.e. they should be OB stars. In the absence of
spectroscopic observations of stars\,1 and 2, we will estimate
their spectral types by using their photometry and assuming that
they are located at the same distance as IRAS\,16455$-$4531.

The distance to IRAS\,16455$-$4531 could be derived from RV of the
strongest peak of the 6.668 GHz methanol maser emission from this
\hii region of $-$91$ \, \kms$ (Walsh et al. 1995; Pestalozzi,
Minier \& Booth 2005). Using the same Galactic constants and the
solar peculiar motion as in Section\,\ref{sec:dist}, one finds the
near and far kinematic distances to IRAS\,16455$-$4531 of
$\approx$5.2 and 9.8 kpc. The former distance agrees with one of
the two likely distances to MN48, which lends some support to the
possibility that MN48 and IRAS\,16455$-$4531 might be associated
with each other. If so, then the difference in RVs of the both
objects would imply that MN48 was ejected from its birth place
with a radial velocity of $\approx$$50 \, \kms$.

Using the extinction law from Rieke \& Lebofsky (1985), the
photometric calibration of optical and infrared magnitudes for
Galactic O stars by Martins \& Plez (2006), and the $J$ and $K$
magnitudes from Table\,\ref{tab:VVV}, we derived the $K$ band
extinction towards stars\,1 and 2 of $A_K$$\approx$2.65 and 1.52
mag, respectively. Then, assuming that both stars are located at 5
kpc, we derived their $K$ band absolute magnitudes of $-$2.84 and
$-$3.63, which correspond to approximate spectral types of B1\,V
and O8\,V, respectively. If subsequent spectroscopy of stars\,1
and 2 will prove that they are indeed massive stars, then the idea
that IRAS\,16455$-$4531 might be the birth place of MN48 would
deserve more thorough consideration.

\subsection{MN48 and the young massive star cluster Westerlund\,1}
\label{sec:wd1}

\begin{figure}
\begin{center}
\includegraphics[width=8cm,angle=0]{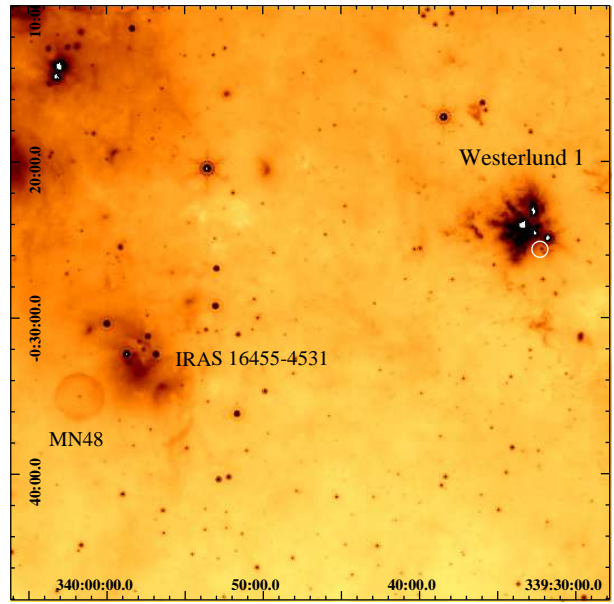}
\end{center}
\caption{{\it Spitzer} MIPS 24\,$\mu$m image of MN48, the \hii
region IRAS\,16455$-$4531 and the \hii region associated with the
young massive star cluster Westerlund\,1. The position of the LBV
W\,243 is indicated by a white circle. The image is oriented with
Galactic longitude (in units of degrees) increasing to the left
and Galactic latitude decreasing upwards. See text for details.}
\label{fig:wd1}
\end{figure}

It is also worthy of note that MN48 is located at only
$\approx$0.5 degree from the young massive star cluster
Westerlund\,1 (see Fig.\,\ref{fig:wd1}). This cluster contains
dozens of massive stars (Clark et al. 2005), including the bona
fide LBV W\,243 (Clark \& Negueruela 2004). Numerous distance
estimates based on various methodologies place this cluster either
in the Crux-Scutum arm (e.g. Kothes \& Dougherty 2007; Koumpia \&
Bonanos 2012) or in the near segment of the Norma arm (e.g. Clark
et al. 2005; Negueruela, Clark \& Ritchie 2010), i.e. the cluster
and MN48 could be at the same distance. The colour excess towards
Westerlund\,1 of $E(B-V)$=4.2$\pm$0.4 mag (Negueruela et al. 2010)
is comparable to that derived for MN48 in Section\,\ref{sec:dist},
which further strengthens the possibility that the distances to
both objects are similar. Taken together, these make Westerlund\,1
a good candidate for the parent cluster to MN48.

The large number of massive stars in the cluster and the high
percentage of binaries among them (Ritchie et al. 2009), imply
that Westerlund\,1 was effective in producing runaway stars
through dynamical few-body encounters at early stages of the
cluster evolution (Gvaramadze \& Gualandris 2011; Fujii \&
Portegies Zwart 2011; Banerjee, Kroupa \& Oh 2012). MN48 could be
one of these stars. At a distance of 5 kpc, MN48 would be
separated from Westerlund\,1 by $\approx$44 pc in projection. This
separation and the age of the cluster of $\sim$5 Myr (Negueruela
et al. 2010) imply a peculiar transverse velocity of MN48 of only
about $10 \, \kms$, provided that the star was ejected soon after
the cluster formation. The actual ejection velocity could be
higher if the star is moving close to our line of
sight\footnote{At the distances of 3.3 and 5 kpc, the systemic
velocity of MN48 of $-$$37 \, \kms$ corresponds to the peculiar
radial velocity of 48 and $13 \, \kms$, respectively.} and/or if
it was ejected because of a binary supernova explosion (in this
case, the kinematic age of MN48 could be much smaller than the age
of the cluster).

The existing proper motion measurements for MN48, however, are
very unreliable, which did not allow us to check the possibility
that this star was ejected from Westerlund\,1. Hopefully, the
space astrometry mission {\it Gaia} will provide proper motion and
parallax measurements accurate enough to identify the parent
cluster to MN48.

\section{Acknowledgements}
This work is supported by the Russian Foundation for Basic
Research grant 16-02-00148. AYK also acknowledges support from the
National Research Foundation (NRF) of South Africa. We thank Dave
Kilkenny for getting for us $V$ and $I_{\rm c}$ band images of
MN48 with the SAAO 1-m telescope and the referee for useful
comments. VVG. This research has made use of observations from the
LCOGT network (programme SAO2015B-002), the NASA/IPAC Infrared
Science Archive, which is operated by the Jet Propulsion
Laboratory, California Institute of Technology, under contract
with the National Aeronautics and Space Administration, the SIMBAD
data base and the VizieR catalogue access tool, both operated at
CDS, Strasbourg, France.

\end{document}